\documentclass[12pt]{article}

\input psfig

\title{
	Statistical Behavior of Finite-Size Partially Equilibrated Systems 
      }

\author{
         {\large Athanasios N. Petridis           } \\
         {\em Department of Physics and Astronomy,}
         {\em Drake University                    } \\
         {\em Des Moines, IA 50311                } \\
         {\em and                                 } \\
         {\em Department of Physics and Astronomy,} 
         {\em Iowa State University               } \\
         {\em Ames, IA 50011                      }
       }

\begin{document}

\maketitle

\begin{abstract}

We examine deviations from Boltzmann-Gibbs statistics for
partially equilibrated systems of finite size. We find that such systems
are characterized by the L\'{e}vy distribution whose non-extensivity parameter 
is related to the number of internally equilibrated subsystems and to
correlations among them. This concept is applied to Quark-Gluon Plasma 
formation.   

\end{abstract}

\vspace{0.2in}

PACS numbers: 05.70.Ln, 05.40.Fb, 25.75.-q

\section{Introduction}

Non-extensive statistics introduced by Tsallis~\cite{Tsallis} has become
a topic of substantial interest and investigation in many areas of physics
including condensed matter, nuclear, and high energy physics. The centerpiece
of such considerations is the power-like L\'{e}vy distribution
\begin{equation}
G_q(x) = C_q \, \left [ 1 - (1 - q)\frac{x}{\lambda_0} 
                \right ]^{\frac{1}{1 - q}},
\label{Levy}
\end{equation}
where $1 \le q < 2$ and $C_q$ is the corresponding normalization constant.
In the limit $q \rightarrow 1$ this reduces to the 
standard Boltzmann-Gibbs exponential factor,
\begin{equation}
g(x) = C \, {\rm exp} \left (-\frac{x}{\lambda_0} \right ).
\end{equation}
In this case the parameter $\lambda_0 \equiv 1/\beta_0$ can be interpreted as 
the temperature of the system in statistical equilibrium. 
(We set the Boltzmann constant equal to unity.)
On the other hand the meaning of Eq.~(\ref{Levy}) is still not 
perfectly clear for $q \ne 1$. In a recent
article Wilk and W\l odarczyk~\cite{Wilk1} have shown that the L\'{e}vy
distribution naturally arises as the mean of the Boltzmann-Gibbs factor
over a Gamma distribution of the temperature parameter, $1/\lambda = 1/T$,
with mean $1/\lambda_0$. In the same article the authors have shown
that the parameter $q$ is determined by the diffusion coefficient and
the dumping constant in the Fokker-Planck equation, i.e, 
the Langevin equation describing gaussian white noise.  
We wish to further elaborate on this observation
by finding other possible origins of this Gamma distribution.

Tsallis' statistics, initially introduced to describe fractal properties
of various systems, maintains the general structure of thermodynamics.
Formal analogies between such statistics and the so-called $q$-oscillators
have been established by means of $q$-calculus and use of 
Jakson derivatives~\cite{Abe}. This statistics has found applications in many 
phenomena including modifications of solar neutrino fluxes~\cite{Kania1},
fluctuations and correlations in high-energy nuclear 
collisions~\cite{Kania1,Wilk2}, the long-flying component of cosmic 
rays~\cite{Wilk3}, chaotic transport in laminar fluid flow~\cite{Solomo},
subrecoil laser cooling~\cite{Bardou} and others. It has also been 
incorporated into fractional quantum statistics~\cite{Kania2}. 

Gross and Votyakov~\cite{Gross} have investigated the statistical 
properties of small systems (systems whose size is comparable with
the range of internal interactions) and have concluded that, in this
case, the grand-canonical distribution is not appropriate and the
thermodynamic limit cannot be taken. Earlier Prosper~\cite{Prosper} had 
already studied temperature fluctuations in finite-size heat baths and
shown that the energy distribution of a thermometer is, in fact,
binomial. It appears that finiteness of the system size and correlations
between its parts due to long range interactions can lead to substantial
deviations from the widely-used Boltzmann-Gibbs distribution. 
An exhaustive coverage of all work related to Tsallis' statistics can be
found in Ref.~\cite{Tsallis2}.
In this article we examine these concepts in the case of partially 
equilibrated systems.

\section{Derivation of the L\'{e}vy Distribution}

First we would like to see how the L\'{e}vy distribution can arise for
finite-size partially equilibrated systems and  
to investigate the meaning of its parameter $q$.  We consider a Gibbs
ensemble of finite-size system replicas each subject to the constraint
of fixed total energy, $E$~\cite{Phillies}. Each system, member of the 
ensemble, is partially equilibrated~\cite{Landau}. It consists of finite-size
subsystems each of which is in internal statistical equilibrium
corresponding to some temperature, $T$. Since the size of the
subsystems and of the whole system is finite the number of subsystems
is also finite. For any member of the ensemble there is a number $k$ of 
subsystems at temperature $T$ (more precisely in the interval between $T$ and
$T+dT$). The ensemble mean of the number of subsystems at temperature
$T$ is $\langle k \rangle$. We define $T_0$, the ensemble mean of the
temperature of all subsystems first averaged over the number of
subsystems of each ensemble member. This is the same as the
temperature average over all the subsystems of all ensemble
members. The more ensemble members have average temperatures close to
$T_0$ or the more subsystems over the ensemble have temperatures close to 
$T_0$ the closer to equilibrium the {\em physical\/} system will
be. We assume that the energy of the whole system is proportional
to the mean temperature,
\begin{equation}
E = c \, T_0,
\end{equation} 
where $c$ is a constant. At the same time the ensemble mean energy
of all subsystems at temperature $T$ will be taken equal to the ensemble mean 
of the number of subsystems at this temperature times the energy of each 
subsystem, i.e.,
\begin{equation}
\langle E (T) \rangle = c \, \langle k \rangle \, T.
\label{MeanE}
\end{equation} 
Under the strong assumption that $E = \langle E(T) \rangle$
we can easily see that
\begin{equation}
\langle k \rangle = \frac{T_0}{T} \equiv \lambda_0 \, \beta,
\label{MeanET}
\end{equation}
where $\lambda_0 = T_0$ and $\beta = 1/T = 1/\lambda$. This condition
correlates the subsystems and may be the result of long-range
interactions within the system. The case $T = T_0$ is 
rather special since, then, Eqs.~(\ref{MeanE},\ref{MeanET}) imply that 
$\langle k \rangle = 1$ and, in this case, there can be only one subsystem 
at temperature equal to the average one. The probability distribution
for encountering exactly $k$ subsystems at temperature $T$ inside one
ensemble member can be considered to be Poissonian for those values of
$k$ reached within one ensemble member,
\begin{equation}
P_k (\beta) = \frac{\langle k \rangle^k {\rm e}^{-\langle k \rangle}}{k!}. 
\label{Poisson}
\end{equation} 
Consequently, the probability for encountering $\alpha$ subsystems
at temperature $T$ within the ensemble is given by the Gamma distribution
\begin{equation}
P_\alpha(\beta) = \frac{\lambda_0 \, (\lambda_0 \beta)^{\alpha -1} \,
                        {\rm e}^{-\lambda_0 \beta}}
                       {\Gamma(\alpha)},
\label{Gamma}
\end{equation}
with first moment $\alpha/\lambda_0$. Upon rescaling $\lambda$ by dividing it
by $\alpha$ and multiplying Eq.~(\ref{Gamma}) by $\alpha$ we obtain the
modified Gamma distribution 
\begin{equation}
f_\alpha \left (\frac{1}{\lambda} \right ) 
	= \frac{\alpha \lambda_0}{\Gamma(\alpha)} \, 
                              \left ( \frac{\alpha \lambda_0}{\lambda}
                              \right )^{\alpha -1} \,
                              {\rm exp} \left (-\frac{\alpha \lambda_0}
                                       {\lambda} \right ),
\label{ModifiedGamma}
\end{equation}
with first moment $1/\lambda_0$. This is the same function as the one 
obtained by Wilk and W\l odarczyk~\cite{Wilk1} starting from the L\'{e}vy
distribution. We may define a parameter $q$ such that
\begin{equation}
\alpha = \frac{1}{q - 1}.
\label{Q}
\end{equation}
Clearly when $q \rightarrow 2$ we have only one subsystem at temperature $T_0$
in the ensemble ($\alpha = 1$). The system is away from equilibrium. We note
that if the first moment of the L\'{e}vy distribution is to be finite then
$q \le 3/2$ so that complete lack of statistical equilibration cannot
be achieved. When $q \rightarrow 1$, $\alpha$ 
tends to infinity; there is an infinite number of subsystems in the ensemble
at $T_0$. In this case the system is equilibrated.
To derive the L\'{e}vy distribution we follow the inverse steps of 
Ref.~\cite{Wilk1}. First we define a variable $\xi$ by means of the equation
\begin{equation}
\xi = \frac{\alpha \lambda_0}{\lambda}.
\end{equation} 
If each subsystem is attributed a Boltzmann-Gibbs factor in a variable
$x$, ${\rm exp}(-x/\lambda)$, then the expectation 
value of this factor over the distribution given by 
Eq.~(\ref{ModifiedGamma}) is
\begin{eqnarray}
\langle {\rm e}^{-x/\lambda} \rangle & = &
 \frac{\alpha \lambda_0}{\Gamma(\alpha)} \,
 \int_{0}^{\infty} {\rm e}^{-x/\lambda} \, \xi^{\alpha - 1}
                               {\rm e}^{-\xi} \, d \left ( \frac{1}{\lambda}
                               \right )  \nonumber \\
 & = & \frac{1}{\Gamma(\alpha)} \int_{0}^{\infty} \xi^{\alpha - 1} \,
         {\rm exp} \left [ -\xi \left ( 1 + \frac{x}{\alpha \lambda_0}
                  \right ) \right ] \, d\xi .
\label{BExpect}
\end{eqnarray}
Even though for a particular ensemble member $1/\lambda$ cannot cover the
entire range $[0, \infty]$ this is cetrainly possible over the ensemble.
We may make the substitution 
$t = \xi \left [1 + x/(\alpha \lambda_0) \right ]$. This yields 
\begin{equation}
\langle {\rm e}^{-x/\lambda} \rangle = 
 \frac{1}{\Gamma(\alpha)} \left ( 1 + \frac{x}{\alpha \lambda_0} 
	\right )^{-\alpha} \int_{0}^{\infty} t^{\alpha - 1} {\rm e}^{-t} dt
	= \left ( 1 + \frac{x}{\alpha \lambda_0} 
	\right )^{-\alpha} 
\end{equation}
and which, by means of Eq.~(\ref{Q}), gives the unnormalized equivalent of
Eq.~(\ref{Levy}). In this derivation the finiteness of the system and
subsystem sizes has been important in order to assume a (discrete) Poissonian 
distribution for $k$.

\section{Entropy Non-Extensivity}

We, now, consider the entropy of the system. Each subsystem being in
internal equilibrium has a Boltzmann-Gibbs-Shannon
entropy,
\begin{equation}
S_i = -\sum_{j=1}^{W_i} p_{ij} \, {\rm ln} \, p_{ij},
\label{OneEntropy}
\end{equation}
where $W_i$ is the number of microstates consistent with a given 
temperature of the subsystem and $p_{ij}$ is the probability for
the $j$th microstate of the subsystem of temperature $T_i$. Since
each subsystem is characterized by a $q$-value, $q_0$, very close to unity
we can apply the approximation $p_{ij}^{q_0 - 1} \approx 1 + (q_0 - 1) \, 
{\rm ln} \, p_{ij}$. In combination with Eq.~(\ref{OneEntropy}) this yields
\begin{equation}
S_i \approx \frac{1}{q_0 - 1} \left [\sum_{j=1}^{W_i}p_{ij} 
                             - \sum_{j=1}^{W_i}p_{ij}^{q_0} \right ]
    = \frac{1}{q_0 - 1} \left [1 - \sum_{j=1}^{W_i}p_{ij}^{q_0} \right ].
\end{equation} 
This power-law entropy was first introduced by Tsallis~\cite{Tsallis3}
and upon optimization under the constraints of normalizability and given 
second moment produces the L\'{e}vy distribution. Even though
this equation was derived as an approximation its applicability is
more general than that of the Boltzmann-Gibbs-Shannon entropy~\cite{Tsallis2}.
This form of the entropy is,
generally, non-extensive as can be seen by calculating the combined entropy
of two stochastically independent (sub)systems,
\begin{equation}
S_{\tau \nu} = \frac{1}{q_0 -1} \left [ 1 - \sum_{l=1}^{W_\tau} 
                                           \sum_{k=1}^{W_\nu}
         p_{\tau l}^{q_0} \, p_{\nu k}^{q_0} \right ]
  = S_\tau + S_\nu - (q_0 - 1) \, S_\tau \, S_\nu ,
\end{equation}
where the combined system is assigned the same $q$-value as its components.
The last equation can be proven by considering the product of the
two individual entropies. Clearly as $q_0 \rightarrow 1$ extensivity
is recovered. If all subsystems are equally close to internal
equilibrium and are stochastically independent the entropy of the whole 
system is 
\begin{equation}
S = \sum_{i} S_i - (q_0 - 1)   \sum_i \sum_{j>i} S_i \, S_j
              - (q_0 - 1)^2 \sum_i \: \: \sum_{j>i} \sum_{k>j>i} S_i \, S_j \,
              S_k - ...
\end{equation}
A microstate of a subsystem is part of a microstate of the system. If a
particular subsystem exists in one ensemble member but does not exist
in another then the contribution to the probability of the subsystem
microstate is zero for the latter member. Therefore, the above summations
extend over all the subsystems that may appear in any ensemble member. We 
observe that in the case of stochastically independent subsystems the total
entropy is smaller than or equal to the sum of the individual entropies,
the equality always holding when the subsystems are internally equilibrated.
However, Eq.~(\ref{MeanET}) induces correlations among subsystems
which remove the stochastic independence. It is due to this effect that
the total entropy of the system is non-extensive even though its parts
are in (or close to) internal equilibrium. This conclusion extends 
the one reached by Landau and Lifshitz~\cite{Landau} on partially 
equilibrated systems. 

In the presence of long-range correlations
and if the stochastic dependence is relatively weak we may approximate
the sum of the $q$-power products of the microstate probabilities of two 
(sub)systems by
\begin{equation}
\sum_{l=1}^{W_\tau} \sum_{k=1}^{W_\nu} p_{\tau l}^{q_0} \, 
                                       p_{\nu k}^{q_0} \approx
  \sum_{\rho=1}^{W_{\tau \nu}} p_{\rho}^{q} + C_{\tau \nu},
\end{equation}
where $W_{\tau \nu}$ is the number of microstates of the combined system, 
$p_{\rho}$ the probability for one such microstate, $C_{\tau \nu}$ is a 
function of $q_0$ that incorporates the correlations and
$q$ is a parameter defined below. We express the functions $C_{\tau \nu}$
in terms of the individual entropy product as
\begin{equation}
C_{\tau \nu} = \kappa \, S_{\tau} \, S_{\nu},
\end{equation}
where $\kappa = \kappa(q_0)$. Using the last two equations we find that the 
total entropy is 
\begin{equation}
S_{\tau \nu} = S_\tau + S_\nu -[q_0 - 1 + \kappa /(q_0 - 1)] \, S_\tau \, S_\nu
             = S_\tau + S_\nu - (q - 1) \, S_\tau \, S_\nu, 
\end{equation}
with $q \equiv q_0 + \kappa /(q_0 - 1)$. When $\kappa \ne 0$ the total
entropy is not extensive even though the individual subsystems may be
close to internal equilibrium. We note that $\kappa$ must be an increasing
function of $q_0 -1$ to prevent the paradoxical situation of $q$ becoming
larger the closer $q_0$ is to unity.

\section{Perspectives for Heavy Ion Physics}

We have seen that the condition of partial statistical equilibration
in a finite-size system leads to a L\'{e}vy distribution that
generalizes the Boltzmann-Gibbs statistics. Long-range correlations
among the internally equilibrated subsystems are important and give
rise to the non-extensivity of the entropy. The parameter $q$ that
characterizes this situation is related to the correlations and
to the number of internally equilibrated subsystems. It is certainly
interesting that the very complicated dynamics of such systems
could be effectively described by a single-parameter statistical
distribution.
One possible application of these ideas is in the formation
of Quark-Gluon Plasma (QGP) in relativistic heavy ion collisions. If such
an unconfined state is actually produced at high temperatures but
relatively low densities then the fragmentation time of the quarks
and gluons to hadrons may be comparable to the relaxation
time of the whole system. In this case the plasma is at most only partially
equilibrated. Thermal photon emission from it will correspond to
a distribution of temperatures and will effectively be described by Tsallis'
statistics that corresponds to the average temperature of the system
and a $q$-value that exceeds 1. It is particularly interesting that
as Walton and Rafelski have shown~\cite{Walton} diffusion of
the charmed quarks in a plasma in equilibrium can be described well only
by the non-extensive Tsallis' statistics. 
On the other hand if the plasma is not 
formed but photons are emitted by a perhaps partially equilibrated hadron 
system the resulting enhansement of photon production (for $q > 1$) as the 
photon transverse momentum increases may mimic photon production from a QGP. 

{\em Acknowledgments:} The author wishes to thank Marshal Luban, Dennis Ross, 
Constantino Tsallis, and Grzegorz Wilk for useful discussions.

\end{document}